\documentclass{aastex}

\usepackage{graphicx}

\usepackage{emulateapj5}
\usepackage{times}

\makeatletter

\newenvironment{inlinefigure}{%
\def\@captype{figure}%
\noindent\begin{minipage}{0.999\linewidth}\begin{center}}
{\end{center}\end{minipage}\smallskip}
\makeatother

\def\keV{ke\kern-0.05emV}
\newcommand{\chandra}{\emph{Chandra}}

\def\asca       {{\em ASCA}\/}
\def\chandra    {{\em Chandra}\/}
\def\rxj1720    {{ RXJ1720.1+2638}\/}

\def\msunyr     {$M_{\odot}\;$yr$^{-1}$}
\def\degd       {$^{\circ}\!$}
\def\second      {{\prime\prime}}

\slugcomment{Accepted for publication in ApJ Letters}

\begin{document}

\title{Evidence for a heated gas bubble inside the ``cooling flow'' region of MKW3s}

\author{P.\ Mazzotta\altaffilmark{1,2}, J.S. Kaastra\altaffilmark{3}, 
F.B. Paerels\altaffilmark{4}, C. Ferrigno\altaffilmark{3},
S. Colafrancesco\altaffilmark{5}, R. Mewe\altaffilmark{3}, and
 W.R. Forman\altaffilmark{1}}
\altaffiltext{1}{Harvard-Smithsonian Center for Astrophysics, 60 Garden St.,
Cambridge, MA 02138; mazzotta@cfa.harvard.edu}
\altaffiltext{2}{ESA Fellow}
\altaffiltext{3}{SRON Laboratory for Space Research Sorbonnelaan
2,3584 CA Utrecht, The Netherlands}
\altaffiltext{4}{Astrophysics Laboratory, Columbia University, 550
West 120th Street, New York, NY 100027, USA}
\altaffiltext{5}{Osservatorio Astronomico di Roma, Via Frascati, 33,
00040, Monteporzio 
Italy}

\shorttitle{Evidence for a heated gas bubble inside the cooling flow region of MKW3s}
\shortauthors{MAZZOTTA ET AL.}

\begin{abstract}
We report on the  deep Chandra observation of central $r=200$~kpc region of
the cluster of galaxies MKW3s which was previously identified as a
moderate cooling flow cluster. The Chandra image reveals two striking
features -- a $100$~kpc long and $21$~kpc wide filament, extending from the 
center to the south-west and
a nearly circular, $50$~kpc diameter depression 90 kpc south of the X-ray
peak.  The temperature map shows that the filamentary structure is colder
while the surface brightness depression is hotter than the average
cluster temperature at any radius. 
The hot and the cold regions indicate that both cooling and
heating processes are taking place in the center of MKW3s.
We argue that the surface 
brightness depression is produced by 
a heated, low-density gas bubble along the line of sight.
We suggest that the heated bubble is produced by
short-lived nuclear outbursts from the central galaxy.
\end{abstract}

\keywords{galaxies: clusters: general --- galaxies: clusters: individual
  (MKW3s) --- X-rays: galaxies --- cooling flows}

\section{Introduction}
It has been argued that 
cooling flows (see Fabian 1994 for a review) are the 
natural state of cluster cores and that more than $40\%$ of  
clusters have flows depositing more than $100$~h$_{50}^{-2}$~M$_\odot$~yr$^{-1}$
(see e.g. Peres et al. 1998).
However, recent   \chandra ~ and XMM results  have not confirmed
cooling flow rates calculated  from
simple models. 
The clearest discrepancy with the ``standard'' cooling flow model
is obtained with the
XMM-RGS:
the spectra from  cooling flow regions in clusters  show a 
remarkable lack of emission lines from gas with $T\le 1$~keV 
 (Peterson et al. 2001, Kaastra et al. 2001, Tamura et al. 2001).
To explain the observed disagreement  with ``simple'' cooling flow
expectations, the above  authors propose a number of possible physical
mechanisms (see also Fabian et al. 2001). 
One mechanism invoked to reduce the cooling rate
is  intracluster medium (ICM) heating by powerful radio sources (mainly AGN) 
in the nucleus of the central galaxy.  
Among the several difficulties of 
such a scenario is that, to date, no cluster shows the ICM heating 
taking place.

We present the temperature map of the central $r=200$~kpc
region of MKW3s derived from a deep \chandra ~ observation.
Even though MKW3s was previously identified as 
a moderate cooling flow cluster  ($\dot M\approx 170$\msunyr, 
Peres et al. 1998), we
argue  that it shows evidence for 
 hot gas bubbles which could be heated 
by AGN activity of the central galaxy.
We also show evidence for a cold filament similar to that seen  
by \chandra ~ in A1795 (Fabian et al. 2000) but we  defer
to a later publication (Mazzotta et al. 2001, in preparation) for 
a discussion.

We use $H_0=50$~km~s$^{-1}$~kpc$^{-1}$, which implies a linear scale of
1.21~kpc per arcsec at the distance of MKW3s ($z=0.045$).

\section{Cluster Observations}

\subsection{X-ray Imaging Analysis}

MKW3s was observed  on April 2000 in  ACIS-I with an exposure of
$\approx 57$~ksec.
We cleaned the data as in Mazzotta et al. (2001).

%%%%%%%%%%%%%%%%%%%%%
% FIGURE 1
%%%%%%%%%%%%%%%%%%%%%

\begin{inlinefigure}
\centerline{\includegraphics[width=0.95\linewidth]{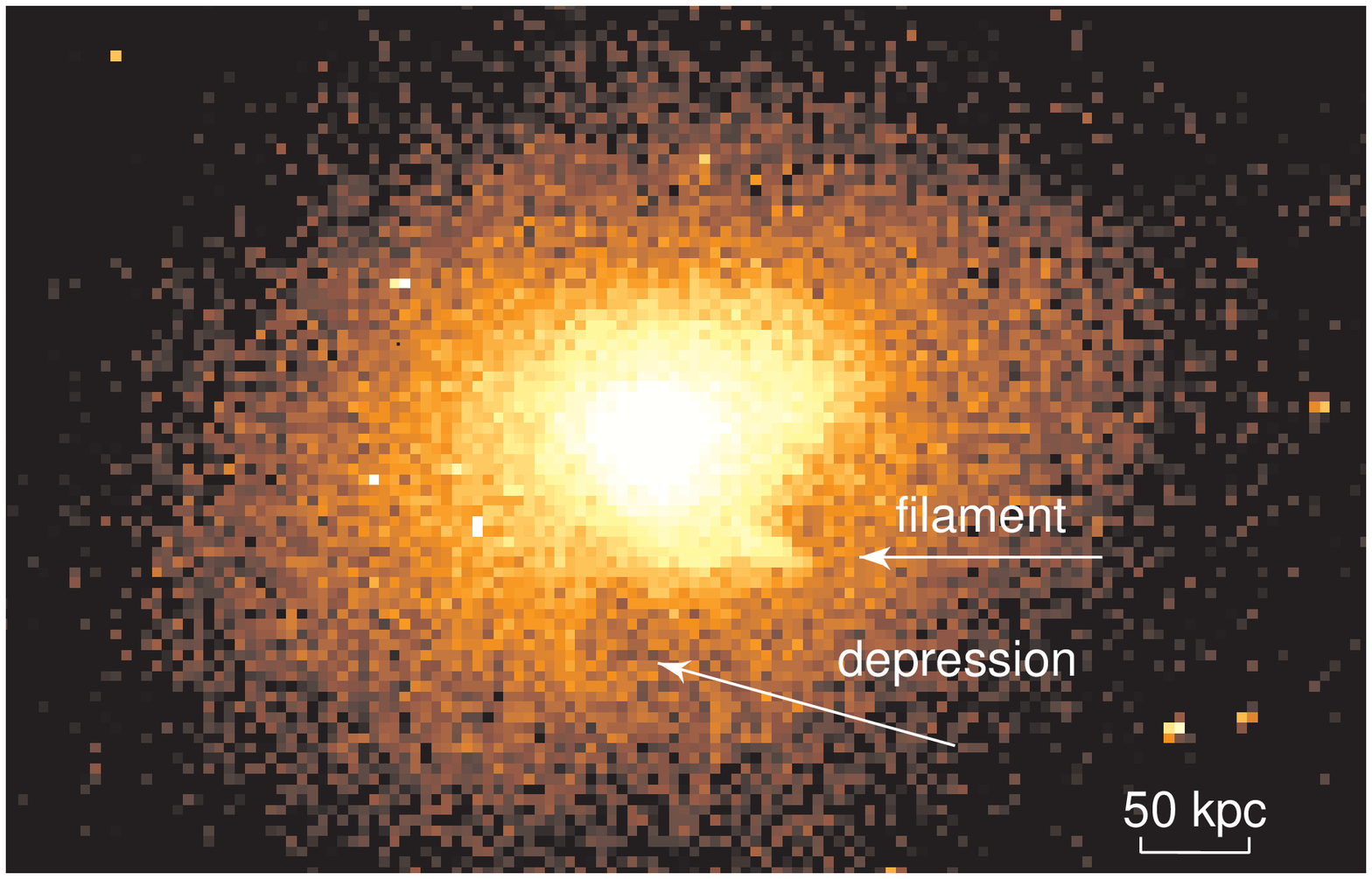}}
\caption{\chandra ~ image of the central $r=200$~kpc region of 
 MKW3s in the $0.5-8$~keV energy band.
Each pixel corresponds to $4^\second \times 4^\second$. 
The arrows indicate the most  prominent features:
the filamentary structure  and the
 surface brightness depression.}\label{fig:x-ray}
\end{inlinefigure}

A background subtracted, vignetting corrected  \chandra ~ image of the 
X-ray emission in the $0.5-8$~keV energy band, from MKW3s, is 
shown in Fig.~\ref{fig:x-ray}. 
The image reveals  
a  surface brightness filament 
extending from the cluster center to the south-west and a 
circular-like depression  south of the X-ray peak.
Both features are evident  in the soft  band ($0.5-2$~keV) as
well as in the hard band ($2-8$~keV) 
indicating that they are produced by a significant  gas density
deviation rather than by  excess  absorption  and/or 
 gas temperature variations.
The surface brightness depression  lies at $\approx 90$~kpc 
($\approx 75^\second$) from the X-ray peak
and exhibits a circular shape with radius $r\approx 25$~kpc 
($r\approx 20^\second)$.
To check the statistical significance of the observed  depression, we
extracted the count rates 
from a circular region inscribed in  the depression  
($r=20^\second$; hereafter region S; see
Fig.~\ref{fig:tmap}) and the  count rates in  two similar 
regions to the east and to the  west of the depression.
The absorbed flux from region S is 
$\approx 14\%$ lower than the flux from the neighboring regions
at a $4.7\sigma$ confidence level.

%%%%%%%%%%%%%%%%%%%%%
% FIGURE 2
%%%%%%%%%%%%%%%%%%%%%

\begin{inlinefigure}
\centerline{\includegraphics[width=0.95\linewidth]{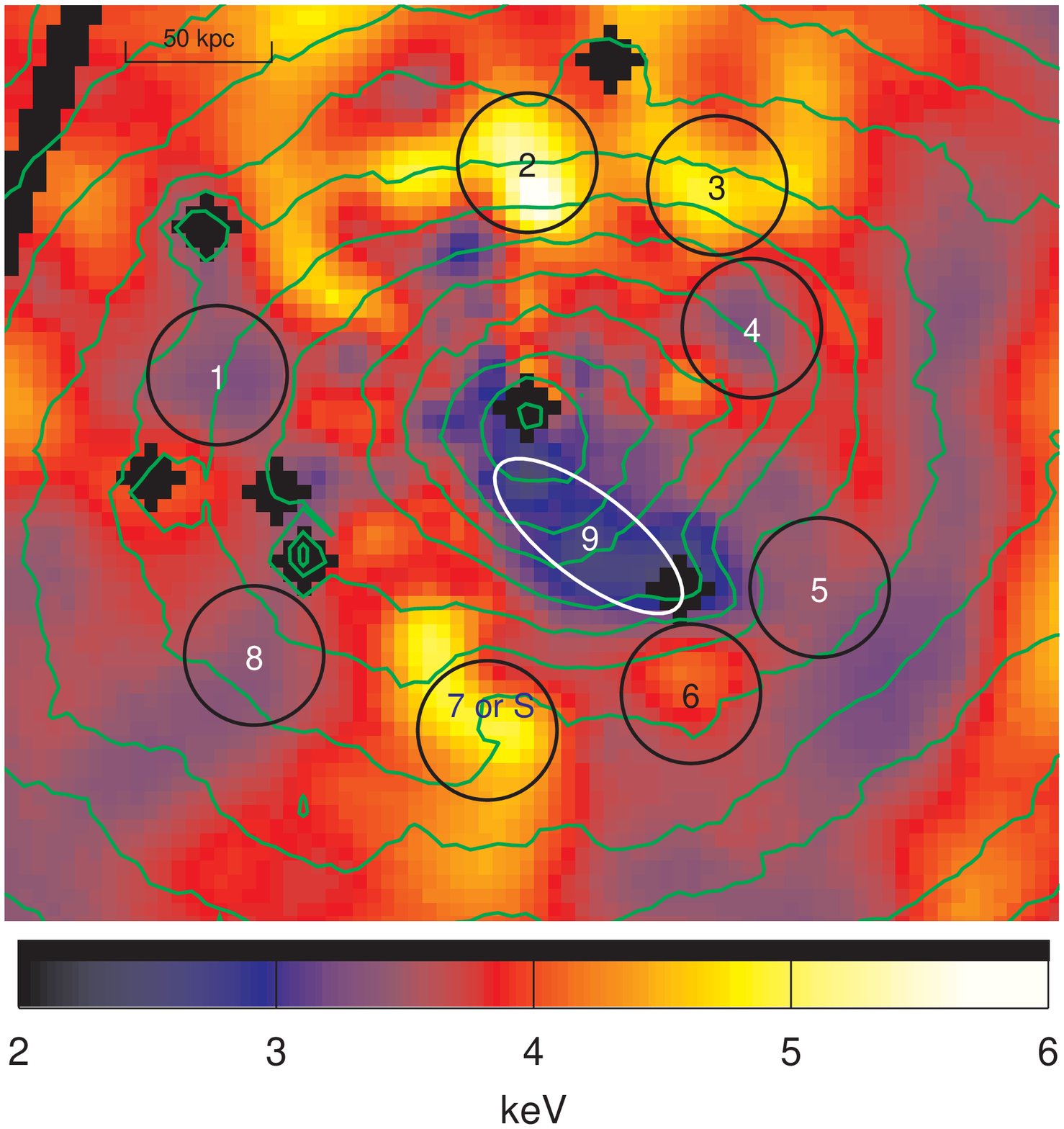}}
\caption{
Temperature map  with overlaid ACIS-I X-ray
surface brightness contours (spaced by a factor of $\sqrt{2}$) in the
0.5-8~keV energy band after adaptive smoothing.   The black cut-out regions identify
  the point sources that were masked out. The  statistical
  error in the temperature map is $<\pm 0.4$~keV at 68\% significance
  level ( $<\pm 0.8$~keV at 90\%). 
The actual temperatures in the regions 
indicated by the cardinal numbers from 1 to 9 are reported in  
Fig.~\ref{fig:regions}.
All regions numbered 1-8 are circles with $r=20^\second$ ($\approx 25$~kpc).} 
  \label{fig:tmap}
\end{inlinefigure}

%
%
%
%%%%%%%%%%%%%%%%%%%%%
% FIGURE 3
%%%%%%%%%%%%%%%%%%%%%

\begin{inlinefigure}
\centerline{\includegraphics[width=0.95\linewidth]{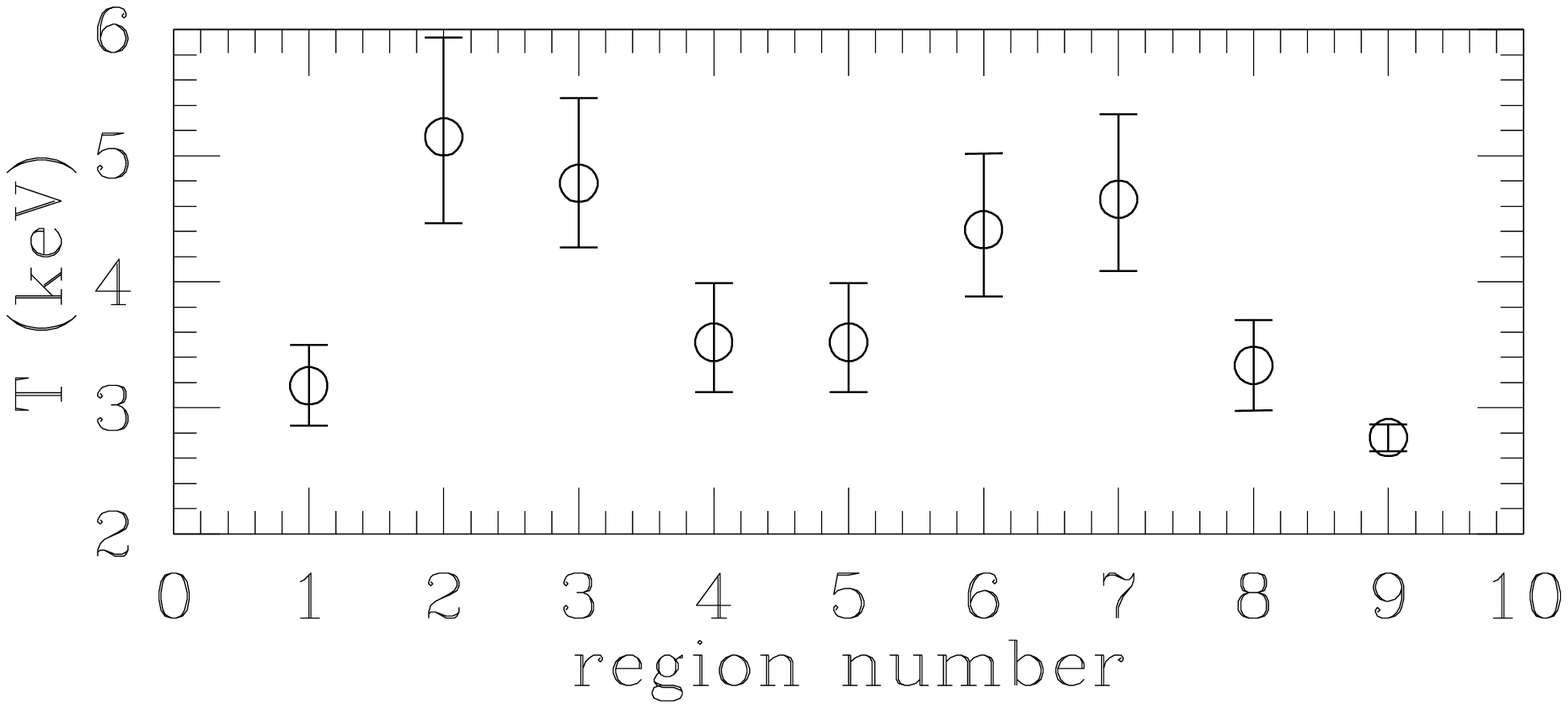}}
\caption{Projected emission-weighted temperature in the
 corresponding regions defined in  Fig.~\ref{fig:tmap} 
(error bars $90\%$ confidence level).}
\label{fig:regions}
\end{inlinefigure}

\subsection{Temperature structure}

All spectral fits were made in the  $0.8-8$~keV energy band.
Moreover, we multiplied the ARFs
by a constant factor 0.93 for $E<2$~keV (Vikhlinin 2000). 
Each spectrum was grouped to have a minimum of 25 counts per bin and 
fitted 
 with an absorbed 
single-temperature thermal (1-T) model (MEKAL).
 We fixed $N_H$ 
to the Galactic value ($N_H=3.04\times 10^{20}$~cm$^{-2}$) and we leave,
as free parameters, the  temperature, the metal abundance, and the emission
measure.

Our technique for temperature map determination is the same as that in
Vikhlinin, Markevitch, \& Murray (2000).
We extracted images in 7 energy bands
($0.80-0.99-2.48-3.50-4.50-6.00-7.01-8.50$~keV)
and smoothed the images with a Gaussian filter
whose width increased  smoothly from
$\sigma=4^\second$ at $r=0^\second$
to  $\sigma=20^\second$ at $r=200^\second$.
We determined the
temperature by fixing
$N_H$ at the galactic value and 
$Z=0.6\ Z_\odot$ (the best-fit  average of the $r=100$~kpc
region).
In Fig.~\ref{fig:tmap} we overlay the  
X-ray contours   
on the  temperature map.
To determine the accuracy of the temperature map, we fit the  
X-ray spectra in 
9 selected regions (indicated in Fig.~\ref{fig:tmap} by 
numbers 1 to 9) 
with a 1-T model as shown in
Fig.~\ref{fig:regions}.

We  derived  the radial temperature profile of the entire cluster 
in twenty elliptical regions centered on the X-ray peak as shown
in  Fig.~\ref{fig:profile}. 
For comparison we report the temperature 
profile derived from an XMM  
and an \asca ~ observation of MKW3s
(Ferrigno et al. 2001, in preparation, Markevitch et al. 1998);  
they are in excellent agreement.

%%%%%%%%%%%%%%%%%%%%%
% FIGURE 4
%%%%%%%%%%%%%%%%%%%%%

\begin{inlinefigure}
\centerline{\includegraphics[width=0.95\linewidth]{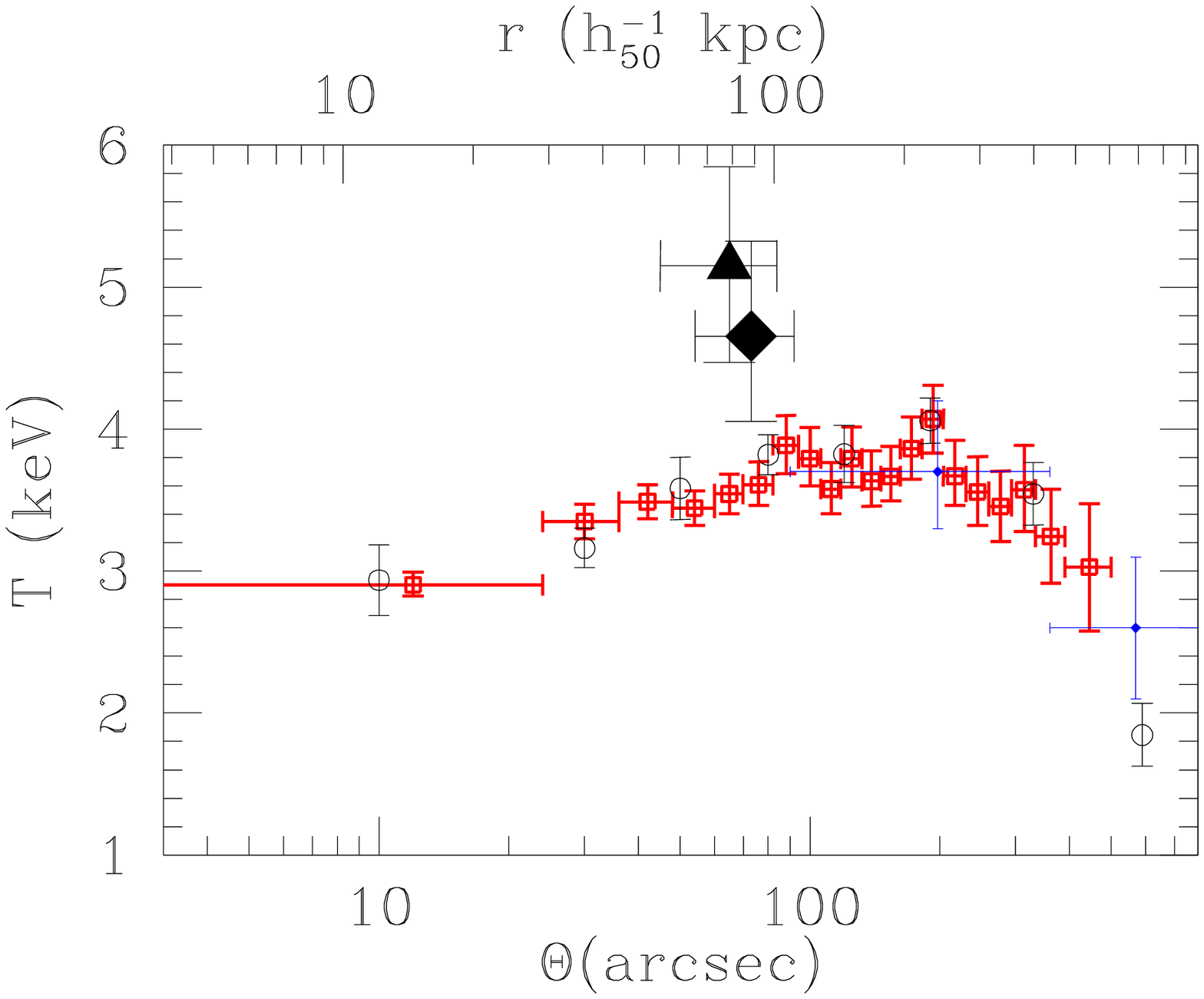}}
  \caption{\chandra ~ and XMM temperature profiles.
Temperatures are extracted from ellipsoidal regions
  centered on the X-ray peak and oriented as the cluster surface
  brightness (error bars $90\%$ confidence level).  The  
ratio of the minor and major axes  is  $\approx 0.7$.
 Red crosses: 
temperature profile obtained from the \chandra ~ observation.
Black circles: temperature profile obtained from the XMM observation.
Blue crosses: temperature measurements obtained from the \asca ~ observation.
Solid diamond and triangle indicate the projected gas temperature of
  the region S (also 7) and 2, respectively (error bars $90\%$ confidence level). }
  \label{fig:profile}
\end{inlinefigure}

\subsubsection{Temperature Structure of region S}

First, we performed a variety of spectral fits to clarify the
nature of the surface brightness depression. By freeing $N_H$  
in the spectral fit for the region S 
we find that it is consistent with the galactic value. Hence, the
  depression is not produced by excess absorption 
but rather  suggests the presence of  a low gas density bubble 
 along the line of sight.
Fig.~\ref{fig:profile} shows that the projected temperature of region S  
(solid diamond)  is higher than
the average temperature at any other radius.

The observed spectrum from region S may  indicate either that
the gas in the bubble  is hotter than the ambient gas or
that the emission from the bubble is dominated by a non-thermal
component (e.g. a power-law).  
To test the origin of the emission from the depression, 
we fit the spectrum using two  simple spectral models:
an absorbed  two-temperature (2-T) model and  an absorbed
1-T plus a power-law model.  
For the first thermal component, we fixed the normalization
to the estimated projected emission measure of the main cluster gas
excluding a sphere of $r_s=25$~kpc located in the plane of the sky
at 90~kpc from the cluster center.
We also fixed $N_H$ to the galactic value and the temperature of the first
thermal component to $T=3.5$~keV, the temperature of the  regions
adjacent to the depression.
For the second component of the 2-T model, we find 
$T=7.5^{+2.5}_{-1.5}$ with
$\chi^2/(d.o.f)= 53.8/(55).$
For the power-law model we find $\Gamma=1.6^{+0.2}_{-0.2}$  $k=1.8\times 10^{-5}$~mJy  with
$\chi^2/(d.o.f) = 56.4/(55)$ (here $S_x=k(E/1$~keV$)^{-\Gamma}$).
Both models fit the observed spectrum very well.

%%%%%%%%%%%%%%%%%%%%%
% FIGURE 5
%%%%%%%%%%%%%%%%%%%%%

\begin{inlinefigure}
\centerline{\includegraphics[width=0.95\linewidth]{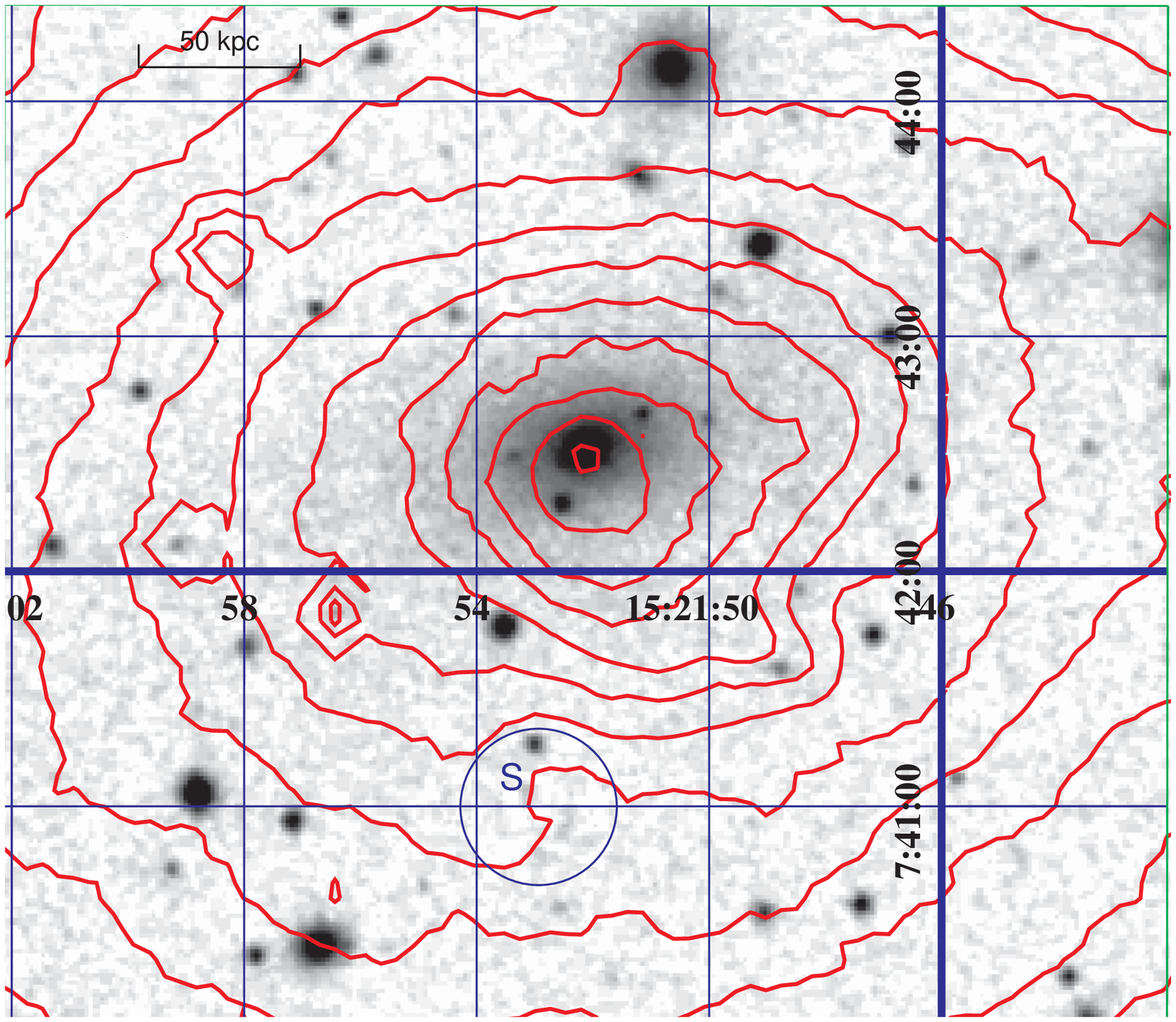}}
\caption{Digitized Sky Survey image with overlaid X-ray
surface brightness contours as in Fig.~\ref{fig:tmap} .  
The circle identifies the region S inscribed in the X-ray surface
brightness depression of Fig.~\ref{fig:x-ray}.} 
  \label{fig:dss}
\end{inlinefigure}

\subsection{Optical and Radio Images}

In Fig.~\ref{fig:dss}  we  overlay the  X-ray
contours  
 on the 
Digital Sky Survey (DSS) image. 
The X-ray brightness peak coincides with the  
optical center of the cluster cD galaxy.
The optical image shows no galaxy concentration
coincident with either the filament or the  X-ray surface
brightness depression.

In  Fig.~\ref{fig:radio} we show
the temperature map  with overlaid  
 VLA radio (1.4~GHz) contours. 
The radio image shows only two bright structures:
the northern radio  source is associated with the  dominant cluster 
galaxy, the origin of the external source to the south is unclear.
This source
has a steep radio spectrum ($\Gamma=3.10$; De Breuck et al. 2000),  
but  steep spectra are common  both  in cluster halos 
(see e.g. Giovannini, Tordi, \& Feretti 1999) as well as in high redshift 
galaxies (see e.g. De Breuck et al.  2000). 
If it is related to the cluster,
it does not coincide with either the X-ray depression or with the
X-ray filament.

\section{Nature of the surface brightness depression}

In this paper we concentrate our  discussion on
 the  nature of the X-ray surface 
brightness depression.
In $\S~2.2.1$ we showed that the
 1-T fit to the spectrum from region S gives a
temperature higher than the radially averaged gas temperature 
at any radius.
This higher temperature may   indicate either that
the gas in the bubble  is hotter than the ambient gas or 
that the bubble is dominated by non-thermal emission.
Unfortunately, simple spectral analysis does not allow us to 
distinguish between these possibilities since  both thermal and
non-thermal
 models 
give acceptable fits to the observed spectrum.
Thus, below we  separately discuss the two possibilities and conclude
 that thermal emission is more likely. 

%
%
%

%%%%%%%%%%%%%%%%%%%%%
% FIGURE 6
%%%%%%%%%%%%%%%%%%%%%

\begin{inlinefigure}
\centerline{\includegraphics[width=0.95\linewidth]{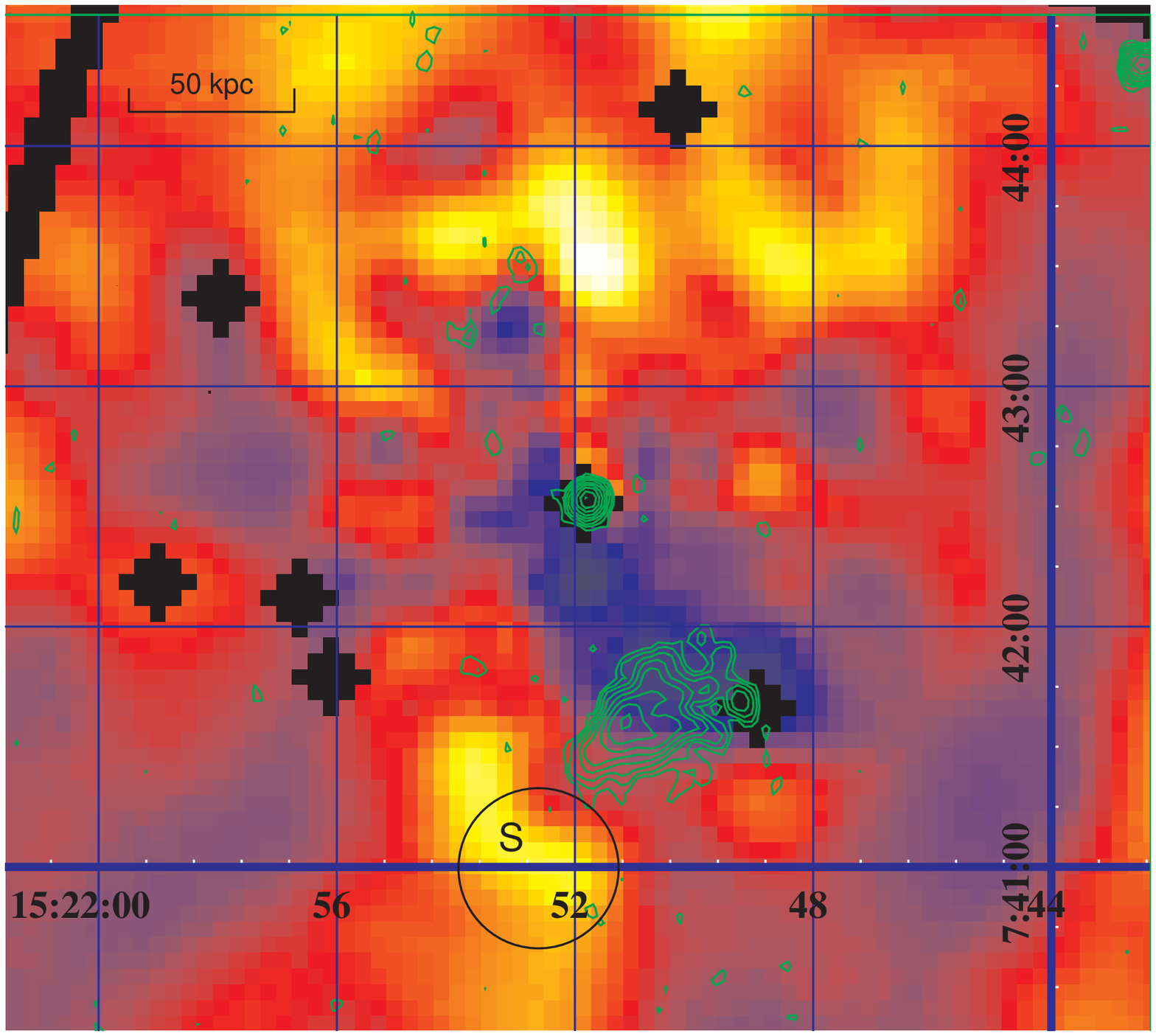}}
  \caption{X-ray temperature map  with  VLA radio (1.4~GHz) contours
  obtained from the First Survey Catalog (contours spaced by a factor of $\sqrt{2}$). The
  circle indicates the  region S.}   
  \label{fig:radio}
\end{inlinefigure}

\subsection{Non-thermal  emission}

If the emission from the bubble is dominated by a power-law component,
then it may be produced 
by  inverse-Compton (IC) scattering of a power-law distribution of
relativistic electrons with the cosmic microwave background (CMB) 
photons (see e.g. 
Harris \& Grindlay 1979).
If this is the case we should also expect 
 synchrotron radio emission.
The radio and the X-ray fluxes provide an estimate
of  the magnetic field strength in the bubble (Harris \& Grindlay 1979). 
 Since the FIRST image of MKW3s  shows no 
significant radio emission associated with the bubble,  at first
we assume that the radio flux $S_r$ is equal to the survey detection
limit: $S_r(1.4$~GHz$)= 1$~mJy. In $\S~2.2.1$ we showed
 that the X-ray spectral fit  
gives $S_x(1$~keV$)=1.8\times 10^{-5}$~mJy and $\Gamma=1.6$.
Using these values we find that  $B_{IC}\approx 8\times 10^{-10}$~Gauss.
Assuming the same radio flux, we estimate the magnetic field strength
using the minimum energy condition (Miley 1980) and we find
 $B_{me}\approx  7\times 10^{-5}$~Gauss.  

The two magnetic field estimates show a great discrepancy as
$B_{me}/B_{IC}\approx 8\times 10^{4}$.
Because $B_{me}\propto S_r^{0.286}$ and $B_{IC}\propto S_r^{0.625}$, 
the above discrepancy is even worse if the actual radio flux is $S_r<1$~mJy.
If we believe  that the minimum  energy condition is
valid (see e.g. Readhead 1994) then it appears unlikely
that the observed X-ray emission from the  bubble can be produced by
IC. Moreover, from the  upper limit on the radio flux, we estimate that the 
synchrotron energy density in the bubble 
is  $<10^{-4}$ times the energy density of
the CMB. Thus, we also can exclude the possibility that the observed
X-ray 
emission is
produced by the
synchrotron self Compton process.

\subsection{Thermal emission} 

The above considerations suggest that the emission from the low 
density bubble is more likely thermal. 
The exact determination of the gas temperature and density of the bubble
depend on the actual shape and on the position of the bubble with
respect to the plane of the sky. 
Let us assume that the bubble is a sphere and is
located close to the plane of the sky. 
In $\S~2.2.1$ we showed that in this case the deprojected gas temperature is
$T_b=7.5_{-1.5}^{+2.5}$~keV.
Using the difference between the flux 
in region S and the adjacent regions, we estimate  the ratio 
$\gamma$ between the  bubble gas density  $n_b$ and the ambient
 gas  density  $n_a$ to be $\gamma=0.71$. 
If the bubble  is at an angle $\theta$ with respect to the plane of
the sky, then the contribution of the 
bubble to the total flux must decrease.
Thus,  to obtain  the
observed depression, the ratio $n_b/n_a$ must be smaller
($\gamma<0.71$).
Consequently the deprojected bubble gas 
temperature should be $T_b>7.5$~keV. 
By requiring  $n_b>0$ ($\gamma>0$) 
we can constrain the position of the bubble to $|\theta|<42$\degd .   
Given the density of the bubble as a function of $\theta$, we estimated
the corresponding deprojected temperature and, thus, the pressure.
By  assuming that the temperature $T_a$ of the ambient gas outside the bubble
is equal to  $T_a\approx 3.5-4$~keV
(see also Fig.~\ref{fig:profile}), we find that, for
$|\theta|<30$\degd ,
the gas pressure  $P_b$ in the  bubble is 1-2 times larger than
the pressure $P_a$ of the surrounding gas.
Unfortunately, for 
 $|\theta|>30$\degd ~ it is not possible to derive an accurate estimate of the 
bubble pressure as its gas  density goes rapidly to zero and the resulting
deprojected temperature is not constrained.
The above considerations strongly
suggest that 
the
gas in the bubble was  heated,  
started to expand  (or is still expanding) and, 
is now  (or soon will  be) in pressure equilibrium with the 
surrounding gas. 

\section{Discussion}

It is not clear how the gas inside the bubble 
can have  been heated to such a high temperature.
Nevertheless, we may expect that the bubble 
formed recently.
If we assume that the bubble expands 
close to the sound speed,  the bubble
age is $t\approx 2.7\times 10^{7}$~yr, 

 much shorter  than the expected age of the cluster.

 The complex temperature structure and the non trivial ellipticity of the
cluster core,  suggest that the gas is not in hydrostatic
equilibrium and/or MKW3s is undergoing a major merger. 
This last interpretation, however,
is not supported by either the optical data
or the X-ray surface brightness analysis which
indicate that MKW3s is a relaxed system. 
Girardi et al. (1997) 
show that MKW3s exhibits no substructure and that
the galaxy velocity distribution is Gaussian and unimodal. 
Moreover the velocity of the cD galaxy is consistent with the cluster mean
and the
velocity dispersion is  consistent with the cluster mean
X-ray temperature. 
Furthermore, the X-ray image does not show evident substructures.
As shown by Buote \& Tsai (1996) the power ratio $P_2/P_0$  of the
multipole terms of order 2 and 0, respectively, represent a critical
indicator of the cluster dynamical status:
for MKW3s, this value  is consistent with those of many relaxed systems 
and significantly lower of those of well known merger clusters.

One possible mechanism for heating gas in the bubble
 is energy injection arising from  activity in the 
nucleus of the cD galaxy. This scenario,  however, seems to be
inconsistent with the  radio map that shows no sign of
strong nuclear activity (see $\S~3$).
However, 
we may be observing a cluster in which 
the black hole  
in the central galaxy undergoes short  intervals ($10^{7}-10^{8}$~yr) of
strong activity followed by periods ($10^{9}-5\times 10^{9}$~yr) 
of relative quiescence (e.g. Binney \& Tabor 1995, Soker et al. 2000).
Such a scenario suggests that the bubble was 
heated during the last radio burst.
If this interpretation is correct and the
gas was heated by radio jets, then the hot regions 
on the north side of the cD galaxy (labeled as regions 2 and 3 in
Fig.~\ref{fig:tmap})  may have been heated
by a northern jet component.
The fact that  we do not detect a significant
corresponding surface brightness depression suggests that the
outburst was  strongly asymmetric. 
Moreover, intermittent energy injection can
contribute to form the observed temperature asymmetry in the core of MKW3s 
(see Churazov et al. 2000 for a discussion of the evolution of buoyant
bubbles in cluster atmospheres).

Finally, the discovery of this heated low-density gas  bubble  
assumes a particular importance as it clearly
shows heating taking place in the cluster core region (previously 
identified as a
cooling flow region) of an apparently ``relaxed'' cluster.
If widespread, such gas heating mechanisms may contribute to
substantially reduce  the cooling rate  and may explain the absence  of 
emission lines from cool ($1$~keV or below) gas in recent cluster 
observations. Furthermore,  we should expect to find them in other
relaxed clusters which would clearly differentiate among the
possibilities for their origin.

\acknowledgments
We thank the referee D. Buote, for useful comments and suggestions.
P.M. thanks D. Harris and M. Markevitch for useful discussions.
P.M. is particularly grateful to A. Vikhlinin for sharing his
software tools that simplified the analysis of this \chandra~ observation. 
P.M. acknowledges an ESA fellowship
 and thanks the Center for
Astrophysics for the hospitality. 
Support for this study was provided
by NASA contract NAS8-39073, grant NAG5-3064, and by the Smithsonian
Institution. 
The Laboratory for Space Research Utrecht is supported
financially by NWO, the Netherlands Organization for Scientific Research.


\begin{references}

%\reference{}  Allen, S.\ W.,  Taylor, G.B.,  Nulsen, P.E.J.,
% Johnstone,  R.M.,  David, L.P.,  Ettori, S., Fabian,  A.C.,  Forman, W.,
%  Jones, C., \& McNamara, B.  2000, \mnras, in press (astro-ph/0101162)


\reference{} Binney, J.\ \& Tabor, G.\ 1995, \mnras, 276, 663 

\reference{} Buote, D.\ A.\ \& Tsai, J.\ C.\ 1996, \apj, 458, 27 

\reference{} Churazov, E.,  Br\"uggen, M., Kaiser, C.R., B\"ohringer,
H.,\& Forman, W. 2000, ApJ in press (astro-ph/0008215)


%\reference{} David, L.\ P., Nulsen, P.\ E.\ J., McNamara,  B.\ R.,  Forman, W., Jones,
%C.,  Ponman, T.,  Robertson, B., \& Wise, M. 2000, ApJ, 
%in press (astro-ph/0010224)

\reference{} De Breuck, C., van Breugel, W., 
R{\"o}ttgering, H.\ J.\ A., \& Miley, G.\ 2000, \aaps, 143, 303 


%\reference{} Ekers, R.\ D.\ \& Simkin, S.\ M.\ 1983, \apj, 265, 85 

\reference{} Fabian, A.\ C.\ 1994, \araa, 32, 277 

\reference {}  Fabian, A.\ C.\,  Sanders, J.S.,  Ettori, S., Taylor,
 G.\ B.,
 Allen, S.\ W.,  Crawford,  C.\ S., Iwasawa, K., \& Johnstone,  R.\ M. 
2000, \mnras, in press (astro-ph/0011547)

\reference{} Fabian, A.\ C., Mushotzky, R.\ F., Nulsen, P.\ E.\ J., \& 
Peterson, J.\ R.\ 2001, \mnras, 321, L20 

%\reference{} Ferrigno et al. 2001, in preparation

\reference{} Giovannini, G., Tordi, M., \& Feretti, L.\ 1999, New 
Astronomy, 4, 141 

\reference{} Girardi, M., Escalera, E., Fadda, D., Giuricin, G., 
Mardirossian, F., \& Mezzetti, M.\ 1997, \apj, 482, 41 



\reference{} Harris, D.\ E.\ \& Grindlay, J.\ E.\ 1979, \mnras, 188, 25 

\reference{} Kaastra, J.\ S., Ferrigno, C., Tamura, T., Paerels, F.\ B.\ 
S., Peterson, J.\ R., \& Mittaz, J.\ P.\ D.\ 2001, \aap, 365, L99 

%\reference{} McNamara, B.\ R., Wise, M.,
% Nulsen, P.\ E.\ J., David, L. P., Sarazin, C. L.,
% Bautz, M., Markevitch, M., Vikhlinin, A.,
% Forman, W.\ R., Jones, C., \& Harris, D.\ E.  2000, \apjl, 534, L135 

\reference{}   Markevitch, M., Forman, W. R.
Sarazin,  C.\ L. \&  Vikhlinin A. ApJ 503, 77, 1998

\reference{} Markevich, M. et al. 2000, CXC memo
(http://asc.harvard.edu/cal ``ACIS'', ``ACIS Background'')

\reference{}  Mazzotta, P.,   Markevitch, M.,
 Vikhlinin, A.,  Forman, W.\ R.,   David, L.\ P., \&
 VanSpeybroeck, L., 2001, ApJ, in press (astro-ph/0102291)

%\reference{} Mazzotta, P. et al., 2001b, in preparation

\reference{} Miley, G.\ 1980, \araa, 18, 165 

\reference{} Peres, C.\ B., Fabian, 
A.\ C., Edge, A.\ C., Allen, S.\ W., Johnstone, R.\ M., \& White, D.\ A.\ 
1998, \mnras, 298, 416 

%\reference{} Peterson, J.\ R., Paerels, F.\ B.\ S., Kaastra, J.\ S., 
%Arnaud, M.,
% Reiprich, T.\ H., Fabian, A.\ C., Mushotzky, R.\ F.,
% Jernigan, J.\ G., \& Sakelliou, I.  2001, \aap, 365, L104 

\reference{} Peterson, J.\ R., et al.
 2001, \aap, 365, L104 

\reference{} Readhead, A.\ C.\ S.\ 1994, \apj, 426, 51 


\reference{} Soker, N., White, R.\ E., David, L.\ P., \& McNamara, B.\ R.\ 
2001, \apj, 549, 832 

\reference{} Tamura, T., et al.
 2001, \aap, 365, L87 


\reference{} Vikhlinin, A., 2000, \chandra ~ calibration memo

\reference{} Vikhlinin, A., Markevitch, M., \& Murray, S.\ S. 2000, \apj,
in press (asro-ph/0008496)

\end{references}
\end{document}